\documentclass[11pt,preprintnumbers,aps,amssymb,nofootinbib,amsmath,superscriptaddress]
{revtex4}
\usepackage{epsfig,epsf}
\usepackage{bm} % puts greek and math symbols in boldface using \bm
\newcommand{\beq}{\begin{equation}}
\newcommand{\beql}[1]{\begin{equation}\label{#1}}
\newcommand{\eeq}{\end{equation}}
\def\bal#1\gal{\begin{align}#1\end{align}}
\newcommand{\ball}[1]{\bal\label{#1}}
\newcommand{\eq}[1]{(\ref{#1})}
\newcommand{\fig}[1]{Fig.~\ref{#1}}
\renewcommand{\sec}[1]{Sec.~\ref{#1}}
\newcounter{topiccounter}
\renewcommand{\b}[1]{{\bm #1}} 
\newcommand{\as}{\alpha_s}

\newcommand{\im}{\,\mathrm{Im}\,}

\newcommand{\aver}[1]{\left\langle #1 \right\rangle}
\begin{document}

\title{Coulomb corrections to photon and dilepton production in high energy pA collisions}

\author{Kirill Tuchin}

\affiliation{
Department of Physics and Astronomy, Iowa State University, Ames, IA, 50011}

\date{\today}


\begin{abstract}
We consider particle production in high energy pA collisions. In addition to the coherent interactions with the nuclear color field,  we take into account coherent interactions with the nuclear electromagnetic Coulomb field. Employing the dipole model, we sum up the  leading multiple color and electromagnetic interactions and derive inclusive cross sections for photon and dilepton production. We found that the Coulomb corrections are up to 10\% at $\sqrt{s}=200$~GeV per nucleon.

\end{abstract}

\maketitle

%%%%%%%%%%%%%%%%%%%%%%%%%%%%%%%%%%%%%%%%
\section{Introduction}

A pivotal feature of high energy pA and AA collisions at RHIC and LHC is large longitudinal coherence length that by far exceeds radii of heavy nuclei. 
In QCD,  color fields of nucleons in a heavy nucleus fuse to create an intense coherent color field, which  has fundamental theoretical and phenomenological importance. Since nuclear force is short-range, only nucleons along the same impact parameter add up to form a coherent field. Because the QCD contribution to the scattering amplitude at high energy  is imaginary, it is proportional to $\as^2$. Thus, the parameter that characterizes the color-coherent field is $\as^2 A^{1/3}\sim 1$, where $A$ is atomic weight. The longitudinal coherence length increases with the collision energy, but decreases with momentum transfer, so that at low energies color coherence  is a non-perturbative phenomenon. At RHIC the longitudinal coherence length is large even for semi-hard transverse momenta (a few GeV's), authorizing application of the perturbation theory to color-coherent processes \cite{Gribov:1984tu,McLerran:1993ka,Kovchegov:1996ty}.

Along with strong color field, heavy-ions also posses strong electromagnetic Coulomb field. The electromagnetic force is long range, so that all $Z$ protons of an ion contribute to the field. Also, the QED contribution to the scattering amplitude is approximately real. As a result, the parameter that characterizes the coherent electromagnetic field is $\alpha Z\sim 1$. Since both parameters $\as^2 A^{1/3}$ and $\alpha Z$ are of the same order of magnitude in heavy ions,  electromagnetic force must be taken into account along with the color one. This observation is a direct consequence of coherence which enhances the electromagnetic contribution by a large factor $Z$. Not all particle production channels in high energy pA and AA collisions are equally affected by the nuclear Coulomb field(s). Our main observation is that gluon emission off a fast quark  is completely unaffected in the eikonal approximation, whereas photon and dilepton production are moderately modified. The central goal of this article is to evaluate the magnitude of the Coulomb corrections to these processes.

The article is structured as follows. In \sec{sec:z} we develop a formalism, inspired by the  Glauber-Mueller  model  \cite{dip} that takes into account both color and electromagnetic coherence by means of  multiple scattering resummation. This formalism is applied in \sec{sec:a}  to calculate the scattering amplitude of color-electric dipole of size $r$ on heavy nucleus. The dipole-nucleus amplitude is employed  in \sec{sec:c} and \sec{sec:d}   to compute inclusive photon  and dilepton cross sections correspondingly. We conclude in  \sec{sec:f} with a discussion  of our results and their  ramifications on  pA and AA phenomenology. 

Strong electromagnetic  interactions in pA and AA collisions were investigated before by many authors \cite{Ivanov:1998dv,Ivanov:1998ka,Ivanov:1998ru,Baltz:2001dp,Lee:2001ea,Bartos:2001jz,Baur:2007zz} who where concerned with pure QED contributions. In this paper we are more interested to study an interplay between the QCD and QED dynamics.

%%%%%%%%%%%%%%%%%%%%%%%%%%%%%%%%%%%%%%%%

\section{Glauber model}\label{sec:z}

 Let the nucleus quantum state be described by the  wave function $\psi_A$ that  depends on positions $\{\b b_a, z_a\}_{a=1}^A$  of all $A$ nucleons, where $\b b_a$ and $z_a$ are the transverse and the longitudinal positions  of a nucleon $a$  correspondingly. (In our notation, transverse vectors are in bold face). If the proton--nucleus scattering amplitude $i\Gamma^{pA}$ is known for a certain distribution of nucleons, then the average scattering amplitude is  
\beql{b9}
\langle \Gamma^{pA}(\b b , s)\rangle\ = \int  \prod_{a=1}^Z d^2\b b_a\, dz_a\, |\psi_A(\b b_1,z_1,\b b_2,z_2,\ldots)|^2 \,\Gamma^{pA}(\b b-\b b_1,z_1,\b b-\b b_2,z_2,\ldots, s)\,.
 \eeq
The scattering amplitude is simply related to the scattering matrix element $S$ as $\Gamma(\b b,s)= 1-S(\b b,s)$. The  later can in turn be represented in terms of the phase shift $\chi$ so that in our case 
\beql{b11}
\Gamma^{pA}( \b b-\b b_1,z_1,\b b-\b b_2,z_2,\ldots, s)=1- \exp\{-i\chi ^{pA}( \b b-\b b_1,z_1,\b b-\b b_2,z_2,\ldots, s)\}\,.
\eeq

At high energies, interaction of the projectile proton with different nucleons  is independent inasmuch as the nucleons do not overlap in the longitudinal direction. This assumption is tantamount to taking into account only two-body interactions, while neglecting the many-body ones \cite{Glauber:1987bb}. In this approximation the phase shift $\chi^{l\bar lZ}$ in the proton--nucleus interaction  is just a sum of the phase shifts $\chi^{l\bar lp}$ in the proton--nucleon interactions and correlations between nucleons in the impact parameter space are neglected. We have
\beql{b13}
\aver{\Gamma^{pA}(\b b, s)}= \aver{1-e^{-i\chi^{pA}}}= 
\aver{1- e^{-i\sum_a \chi^{pN}}}= 1- e^{-i\sum_a \aver{\chi^{pN}}}\,,
\eeq
where in the last term $\aver{\ldots}$ stands for an average over a single nucleon 
position in the nucleus, defined below in  \eq{b22}. 
To the leading order in coupling $\as$, the phase shift $\chi^{pN}$ can be expanded as 
$-i\chi^{pN}= \ln (1-\Gamma^{pN})\approx -\Gamma^{pN}$.  Therefore, we can write
\beql{b15}
 \aver{\Gamma^{pA}(\b b, s)}=1- \exp\Big\{-\sum_a\langle \Gamma^{pN}(\b b,s)\rangle\Big\}.
\eeq

Strong and electromagnetic contributions decouple in the elastic  scattering amplitude at the leading order in respective couplings:
\beql{b17} 
\Gamma^{pN}= \Gamma^{pN}_\text{s}+\Gamma^{pN}_\text{em}\,.
\eeq
Indeed, as we discuss below $i\Gamma^{pN}_\text{em}$ is real, while $i\Gamma^{pN}_\text{s}$ is imaginary, which is a consequence of the fact that $SU(3)$ generators are traceless. Owing to \eq{b17} we can cast \eq{b15} in the form  
\beql{b19}
 \aver{\Gamma^{pA}(\b b, s)}=1- \exp\Big\{- A \aver{\Gamma^{pN}_\text{s}(\b b,s)} -Z\aver{\Gamma^{pN}_\text{em}(\b b,s)  }\Big\}\,,
\eeq
where $Z$ is the number of protons.  In the Glauber model we  average over the nucleus using the nuclear density $\rho$ as follows
\beql{b22}
\aver{\Gamma^{pN}_\text{s}(\b b,s)}= \frac{1}{A}\int_{-\infty}^\infty dz_a \int d^2b_a\, \rho(\b b_a,z_a)  \Gamma^{pN}_\text{s}(\b b-\b b_a,s)\,.
\eeq 
Neglecting the diffusion region,  nuclear density is approximately constant  $\rho = A/(\frac{4}{3}\pi R_A^3)$ for points inside the nucleus and zero otherwise. The range of the nuclear force is about a fm, which is much smaller than the radius $R_A$ of a heavy nucleus. Therefore, $\b b\approx \b b_a$ and 
\beql{b25}
\aver{\Gamma^{pN}_\text{s}(\b b,s)}= \frac{1}{A}\, 2\sqrt{R_A^2-b^2}\, \pi R_A^2\, \rho \Gamma^{pN}_\text{s}(0,s)\,.
\eeq
In this approximation the total proton-nucleon cross section is $\sigma^{pN}(s)= 2\pi R_p^2 \Gamma^{pN}_\text{s}(0,s)$, with $R_p$ being proton's radius, so that
\beql{b27}
\aver{\Gamma^{pN}_\text{s}(\b b,s)}= \frac{1}{A}\rho T(b)\frac{1}{2}\sigma^{pN}(s)\,, 
\eeq
where $T(b)=2\sqrt{R_A^2-b^2}$ is the thickness function. It follows from \eq{b27} that 
$A\langle\Gamma^{pN}_\text{s}\rangle\sim \as^2 A^{1/3}$, which implies that \eq{b19} sums up  terms of order $\as^2 A^{1/3}\sim 1$ at $\as\ll 1$. Indeed, the leading strong-interaction contribution to the $pN$ elastic scattering amplitude  corresponds to double-gluon exchange. Note also, that  the corresponding  $\langle i\Gamma^{pN}_\text{s}\rangle$ is purely imaginary.  

Proton density in the nucleus is $Z\rho /A$, hence
\bal\label{b29}
\aver{\Gamma^{pN}_\text{em}(\b b,s)}=& \frac{1}{Z}\int_{-\infty}^\infty dz_a \int d^2b_a\, \frac{Z}{A}\rho(\b b_a,z_a)  \Gamma^{pN}_\text{em}(\b b-\b b_a,s)\\
=
&\frac{1}{A}\rho \int d^2b_a \, T(b_a) \Gamma^{pN}_\text{em}(\b b-\b b_a,s)\,.
\gal
Electromagnetic interaction is long-range, therefore all values of impact parameter $b$ contribute to the total cross section. Moreover, the leading logarithmic contribution comes from impact parameters far away from the nucleus $b\gg b_a\sim R_A$. In this case,
\beql{b31}
\aver{\Gamma^{pN}_\text{em}(\b b,s)}= \frac{1}{A}\rho \, \Gamma^{pN}_\text{em}(\b b,s)\int d^2b_a 2\sqrt{R_A^2-b_a^2}=\Gamma^{pN}_\text{em}(\b b,s)\,,\quad b\gg R_A \,.
\eeq
However, if we are interested in differential cross section at  impact parameters $b\sim R_A$ no such approximation is possible. The leading electromagnetic contribution to elastic  $pN$ scattering amplitude arises from one photon exchange; the corresponding $\langle i\Gamma^{pN}_\text{em}\rangle$ is purely real. We note, that \eq{b22} sums up terms of order $\alpha Z\sim 1$ at $\alpha\ll 1$.

The total $pA$ cross section can be computed using the optical theorem as follows
\beql{b33}
\sigma_\text{tot}^{pA}(s)= 2\int d^2b \im[i\Gamma^{pA}(\b b,s)]= 
2\int d^2b\left\{ 1-\exp[-A \aver{\Gamma^{pN}_\text{s}(\b b,s)}]\cos[Z\aver{i\Gamma^{pN}_\text{em}(\b b,s)}]\right\}\,.
\eeq

%%%%%%%%%%%%%%%%%%%%%%%%%%%%%%%%%%%
\section{Dipole-nucleus scattering}\label{sec:a}

Similarly to the proton--nucleus scattering, one can consider scattering of  color \emph{and} electric singlet  $q\bar q$ pair  (dipole) of size $\b r$ off a heavy nucleus. 
Since a single gluon exchange is an inelastic process, the leading in $\as$ contribution to the elastic scattering amplitude  comes from the  double gluon exchange given by
\beql{b37}
A\aver{\Gamma^{q\bar qN}_\text{s}(\b b,s;\b r)}=\frac{2C_F}{N_c}\rho T(b)\frac{1}{2}\,\pi r^2\, \as^2\ln \frac{1}{r\mu}\,,
\eeq
where $\mu$ is an infrared scale and $s$ the center-of-mass energy squared,  while the leading in $\alpha$ term arises from a singe photon exchange given by 
\beql{b39}
Z\aver{i\Gamma^{q\bar qN}_\text{em}(\b b,s;\b r)}= \frac{Z}{A}\rho\, 2 \alpha  \int d^2b_a\, T(b_a)\,\ln 
\frac{|\b b-\b b_a-\b r/2|}{|\b b-\b b_a+\b r/2|}\,.
\eeq

In this article we employ a simple but quite  accurate ``cylindrical nucleus" model (see e.g.\ \cite{Kovchegov:2001sc,Kharzeev:2004yx}). Namely, we  set  $T(b)=2R_A$ if $b<R_A$ and  zero otherwise.  The impact parameter integrals  in \eq{b33},\eq{b39} can now be taken exactly. In particular, integration over $\b b_a$ is described in Appendix. Since in QCD $r\ll R_A$ we can neglect a very narrow region $|\b b- \b r/2|< R_A<|\b b+\b r/2|$ in which case \eq{app5} yields for the electromagnetic term in the elastic dipole--nucleon scattering amplitude
\ball{b40}
\aver{i\Gamma^{q\bar qN}_\text{em}(\b b,s;\b r)}=2\alpha \left[ -\frac{\b b\cdot \b r}{R_A^2}\theta(R_A-b) + \ln \frac{|\b b-\b r/2|}{|\b b+\b r/2|}\theta(b-R_A)\right]\,.
\gal
The total cross section for dipole-nucleus scattering has the same form as  \eq{b33} and can be now written as 
\ball{b41}
\sigma_\text{tot}^{q\bar qA}(s;r)= &
2\int d^2b\left\{ 1-\exp\left[-A \aver{\Gamma^{q\bar qN}_\text{s}(\b 0,s;\b r)}\right] \cos\left( 2\alpha Z \frac{\b b\cdot \b r}{R_A^2}\right)\right\}\theta(R_A-b)\nonumber\\
&+2\int d^2b\left\{ 1-\cos\left( 2\alpha Z \ln \frac{|\b b-\b r/2|}{|\b b+\b r/2|}\right)\right\}\theta(b-R_A)\,.
\gal
In the first line of \eq{b41} we can replace the cosine by one, because $r\ll R_A$, $\alpha Z\sim 1$. The corresponding contribution to the cross section is
\beql{b43}
\sigma_\text{s}^{q\bar qA}(s;r)= 2\pi R_A^2 \left\{ 1-\exp[-A \aver{\Gamma^{q\bar qN}_\text{s}(0,s;\b r)}]\right\}\,,
\eeq
which is a purely QCD term, hence the subscript ``s"  for the ``strong" interaction. Integral in the second line of \eq{b41} can be taken exactly and yields the QED contribution \cite{Kopeliovich:2001dz,Tuchin:2009sg}
\ball{b44} 
\sigma_\text{em}^{q\bar qA}(s;r)&\equiv 2\int_{R_A}^{b_\text{max}} db\, b \int_0^{2\pi} d\phi  \left\{ 1- \cos\left(\alpha Z\ln 
\frac{b^2+r^2/4-br\cos\phi}{b^2+r^2/4+br\cos\phi}\right)\right\} \nonumber\\
&= 4\pi r^2 (\alpha Z)^2\ln\frac{b_\text{max}}{R_A}
= 4\pi r^2 (\alpha Z)^2 \ln\frac{s}{4m_q^2m_NR_A}\,,
\gal
where $m_q$ and $m_N$ are quark and nucleon masses correspondingly. Terms of order $r^2/R_A^2$ are neglected in \eq{b44}.\footnote{I would like to stress that approximation $r\ll R_A$  holds only if the dipole size $r$ is determined by a QCD scale. For example, in photon emission $r\sim 1/k$, where $k$ is photon's momentum, in dilepton production $r\sim 1/M$, where $M$ is dilepton's invariant mass. Therefore, only if $k$ and $M$ are at or above $\sim 1/R_p\sim 200$~MeV can this approximation be used. The original calculation of \cite{Bethe:1934za,Bethe:1954zz} was done in QED in the opposite limit of a point-like nucleus $r\gg R_A$.} Energy dependence arises from the  long distance cutoff $b_\text{max}= s/(4m_Nm_q^2)$ of the $b$-integral.

The total cross section is thus simply a sum of the QCD and QED terms
\ball{b46}
\sigma_\text{tot}^{q\bar qA}(s;r)=\sigma_\text{s}^{q\bar qA}(s;r)+ \sigma_\text{em}^{q\bar qA}(s;r)\,.
\gal 
 From comparison of \eq{b43} and \eq{b44} it is clear that the QED contribution to the total cross section is suppressed relative to the QCD term by $(r\alpha Z/R_A)^2\log s$, hence the Coulomb correction is largest for soft processes with larger $r$.  Since the largest dipole size is of order  $R_p$, the smallest suppression factor is of order $(\alpha Z)^2/A^{2/3}\log s$,  which for gold nucleus is about $0.1$ at $\sqrt{s}=200$~GeV. Because, $Z\sim A$, the relative contribution of the Coulomb correction increases with $A$.

At high energies,  dipole--nucleon scattering amplitude acquires energy dependence $\Gamma^{q\bar qN}\sim s^{1+\Delta}$, where to the leading order in QCD $\Delta_\text{s}=4\ln 2(\as N_c/\pi)$ \cite{Balitsky:1978ic,Kuraev:1977fs} and in QED  $\Delta_\text{em}= (11/32) \pi \alpha^2$ \cite{Gribov:1970ik,Mueller:1988ju}. Since $\Delta_\text{em}\ll \Delta_\text{s}$ we can neglect the effect of QED evolution. A phenomenological way to take QCD evolution into account 
is to  parameterize the scattering amplitude in terms of quark saturation momentum $\tilde Q_s$ and  anomalous dimension $\gamma$ as follows 
\beql{b53}
A\aver{\Gamma^{q\bar qN}_\text{s}(0,s;\b r)}=\frac{1}{4}(r^2 \tilde Q_s^2)^\gamma\,,
\eeq
where $\tilde Q_s^2\approx 0.16 A^{1/3}$~GeV$^2$ and $\gamma\approx 0.63$ \cite{Mueller:2002zm}. Numerical value of the saturation momentum is known from DIS and heavy-ion phenomenology (see e.g.\ \cite{Kovchegov:2012mbw}).

%%%%%%%%%%%%%%%%%%%%%%%%%%%%%%%%%
\section{Photon production}\label{sec:c}

In this and the next section we discuss photon and dilepton production in high energy pA collisions. Photon production without electromagnetic corrections was calculated in \cite{JalilianMarian:2005zw}. If we assume the validity of the collinear factorization on the proton side, the problem reduces to computing the photon and dilepton production in qA collisions.\footnote{One should be cautious with the collinear factorization of dilute projectiles at high energies since it is not valid  in exclusive processes, see e.g.\ \cite{Li:2008se}, and is violated  even in some inclusive processes  \cite{Tuchin:2012cd,Altinoluk:2011qy}.}  We adopt  the following notations: four-momenta of incoming quark, photon and outgoing quark are $q$, $k_1$ and $k_2$ correspondingly; bold face denotes their respective  transverse components; $z=k_{1+}/q_+$. Transverse coordinates of incoming quark, photon and outgoing quark in the amplitude are $\b u$, $\b x_1$ and $\b x_2$; those in the complex conjugated amplitude are distinguished by a prime; $\b r= \b x_1-\b x_2$, $\b r'=\b x_1'-\b x_2'$, $\b b= (\b x_1+\b x_2)/2$, $\b b'= (\b x'_1+\b x'_2)/2$. 
We also define the following scattering matrix element 
\beql{b35}
\mathcal{S}(\b b,\b r)= 1-\im [i\Gamma^{q\bar qA}(\b b,s;\b r)]= \exp[-A \aver{\Gamma^{q\bar qN}_\text{s}(\b b,s;\b r)}]\cos[Z\aver{i\Gamma^{q\bar qN}_\text{em}(\b b,s;\b r)}]\,.
\eeq

With these notations we can write down the double-inclusive cross section as follows 
\cite{Gelis:2002ki,Baier:2004tj,Dominguez:2011wm}
\bal\label{c9}
\frac{d\sigma^{qA\to \gamma q X}}{d^2 k_1 d^2k_2 dz}=& \frac{1}{2(2\pi)^5}\int d^2u\, d^2 u'\,  d^2x_1\, d^2x_2\,  d^2x_1'\, d^2x_2'\,
e^{-\b k_1\cdot (\b x_1-\b x_1')-i\b k_2\cdot (\b x_2-\b x_2')}\nonumber\\
&
\times \phi^{q\to q\gamma}(\b r,\b r',z)
\big[ -\mathcal{S}((\b x_2'+\b u)/2,\b x_2'-\b u)-\mathcal{S}((\b x_2+\b u')/2,\b x_2-\b u')
\nonumber\\
&
+\mathcal{S}((\b x_2'+\b x_2)/2, \b x_2'-\b x_2)+\mathcal{S}((\b u+\b u')/2,\b u-\b u')\big]\,,
\gal
where the square of the light-cone wave-function 
\beql{c11}
\phi^{q\to q\gamma}(\b r,\b r',z)= \frac{2e_f^2}{(2\pi)^2}\frac{\b r\cdot \b r'}{r^2r'^2}\frac{1+(1-z)^2}{z}\delta(\b u-z\b x_1-(1-z)\b x_2)\delta (\b u'-z\b x_1'-(1-z)\b x_2')
\eeq
describes photon emission off quark in the chiral limit.  According to \eq{b35},\eq{b27},\eq{b29} we have
\beql{c15}
\mathcal{S}(\b b,\b r)=\exp\left\{-\frac{2C_F}{N_c}\rho T(b)\frac{1}{2}\,\pi r^2\, \as^2\ln \frac{1}{r\mu}\right\}\cos\left\{\frac{Z}{A}\rho\, 2 \alpha  \int d^2b_a\, T(b_a)\,\ln 
\frac{|\b b-\b r/2-\b b_a|}{|\b b+\b r/2-\b b_a|}\right\}\,.
\eeq
Integration over the final quark transverse momentum $\b k_2$ gives the single-inclusive cross section
\ball{c19}
\frac{d\sigma^{qA\to \gamma q X}}{d^2 k_1 dz}&= 2\alpha e_f^2\frac{1+(1-z)^2}{z}\int d^2\tilde b\int \frac{d^2r}{(2\pi)^2}\int \frac{d^2r'}{(2\pi)^2}e^{-i\b k_1\cdot (\b r-\b r')}
\frac{\b r\cdot \b r'}{r^2 r'^2}\,\phi^{q\to q\gamma}(\b r,\b r',z)\nonumber\\
& \times \left[-\mathcal{S}(\tilde {\b  b}+z\b r/2, z\b r)-\mathcal{S}( \tilde {\b  b}+z\b r'/2,z\b r')+1 +\mathcal{S}(\tilde {\b  b}+z(\b r+\b r')/2, z(\b r-\b r'))\right]\,.
\gal 
where  $\tilde {\b  b}= \b b- \b r/2$.

As in the previous section we utilize the ``cylindrical nucleus" model to take the impact parameter integrals. In particular, taking integral over $\b b_a$ in \eq{c15} yields
\ball{c21}
\mathcal{S}(\b b, \b r )&=e^{-\frac{1}{4}(\tilde Q_s^2r^2)^\gamma}\theta(R_A-b)+ 
\cos\left(2 \alpha Z \,\ln 
\frac{|\b b-\b r/2|}{|\b b+\b r/2|}\right)\theta(b-R_A)
\gal
up to terms of order $r^2/R_A^2$. With the same accuracy, integration over $\b b$ in \eq{c19} can now be done explicitly using \eq{b33},\eq{b35},\eq{b43},\eq{b44}:
\ball{c23}
\frac{d\sigma^{qA\to \gamma q X}}{d^2 k_1 dz}&= \alpha e_f^2\frac{1+(1-z)^2}{z}\int \frac{d^2r}{(2\pi)^2}\int \frac{d^2r'}{(2\pi)^2}e^{-i\b k_1\cdot (\b r-\b r')}
\frac{\b r\cdot \b r'}{r^2 r'^2}\,\phi^{q\to q\gamma}(\b r,\b r',z)\nonumber\\
& \times \left[\sigma_\text{tot}^{q\bar qA}(s; z\b r)+\sigma_\text{tot}^{q\bar qA}(s;z\b r')-\sigma_\text{tot}^{q\bar qA}(s; z(\b r-\b r'))\right]\,.
\gal 
Eq.~\eq{c23} can be cast  into a factorized form by employing  the following identities  \cite{Kovchegov:2001sc,Kharzeev:2003wz}
\bal
&\int d^2x \, e^{-i\b k\cdot \b x}\frac{\b x}{x^2}= -2\pi i\frac{\b k}{k^2}\,,\label{c25}\\
&\int d^2x'\frac{\b x'\cdot (\b x+\b x')}{\b x'^2(\b x+\b x')^2}= \pi\ln \frac{1}{x^2}\,,\label{c26}
\gal
The result reads
\ball{c27}
\frac{d\sigma^{qA\to \gamma q X}}{d^2 k_1dz}&= \frac{\alpha}{(2\pi)^3} e_f^2\frac{1}{k_1^2}\frac{1+(1-z)^2}{z}\int d^2x\,e^{-i\b k_1\cdot \b x}\,\ln \frac{1}{ x\mu}\,\b\nabla^2_{\b x}\sigma_\text{tot}^{q\bar qA}(s;z\b x)\,.
\gal
The electromagnetic contribution can be calculated exactly:
\ball{c31}
\frac{d\sigma^{qA\to \gamma q X}_\text{em}}{d^2 k_1dz}&= \frac{\alpha}{(2\pi)^2} e_f^2\frac{8\pi}{k_1^4}\frac{1+(1-z)^2}{z}\,4\pi z^2(\alpha Z)^2\ln\frac{s}{4m_q^2m_NR_A}\,,
\gal
where we used 
\ball{c29}
\int d^2x\,\ln \frac{1}{x}\, e^{-i\b k\cdot \b x}= \frac{2\pi }{ k^2}\,.
\gal

To obtain a qualitative estimate of the QCD contribution to the inclusive cross section \eq{c27},  note that unless $x<2/k_1$ the exponent is rapidly oscillating. Furthermore, integrand is exponentially suppressed at $zx>2/\tilde Q_s$. We thus obtain 
\ball{c39}
\frac{d\sigma^{qA\to \gamma q X}_\text{s}}{d^2 k_1 dz}&\approx \frac{\alpha}{(2\pi)^2} e_f^2\frac{1}{k_1^2}\frac{1+(1-z)^2}{z} \int_0^{x_0}dx\,\ln \frac{1}{x}\,\partial_x[x\partial_x(\sigma_\text{tot}^{q\bar qA}(s;zx))]\,,
\gal
where $x_0$ is the smallest of three scales $2/k_1$, $2/(z\tilde Q_s)$ and $1/\mu$. Expanding \eq{b43} with \eq{b53} at small $zx$ we  find
\ball{c41}
\sigma_\text{s}^{q\bar qA}(s;zx)\approx  2\pi R_A^2\frac{1}{4} (\tilde Q_s^2z^2x^2)^{\gamma}\,, 
\gal
which upon substitution into \eq{c39} and combining with \eq{c31} produces 
\ball{c43}
\frac{d\sigma^{qA\to \gamma q X}}{d^2 k_1 dz}\approx &\frac{\alpha}{\pi} e_f^2\frac{1}{k_1^4}\frac{1+(1-z)^2}{z}\,  \left[ 8\pi z^2(\alpha Z)^2 \ln\frac{s}{4m_q^2m_NR_A}+\frac{1}{4}\gamma k_1^2 R_A^2\tilde Q_s^{2\gamma} (x_0z)^{2\gamma}\ln \frac{1}{x_0} \right]\,.
\gal
We see that the ratio of the QCD and the QED terms is of order $ (R_A\tilde Q_s)^2(k_1^2/\tilde Q_s^2)^\eta$, with $\eta =1$, if $k_1\ll \tilde Q_s$ and $\eta= 1-\gamma$, if $k_1\gg \tilde Q_s$. 
Thus, the QED interactions have the largest relative impact  at small photon transverse momenta and in more peripheral events. Note also that the role of QED interactions diminishes with energy because  the saturation momentum increases  as a power of energy, whereas the QED contribution is only logarithmic. The distinct feature of $z$-dependence of inclusive photon production cross section is that it vanishes in the eikonal limit  $z\to 0$, which is evident from  \eq{c19}. 

%%%%
\begin{figure}[ht]
     \includegraphics[width=8cm]{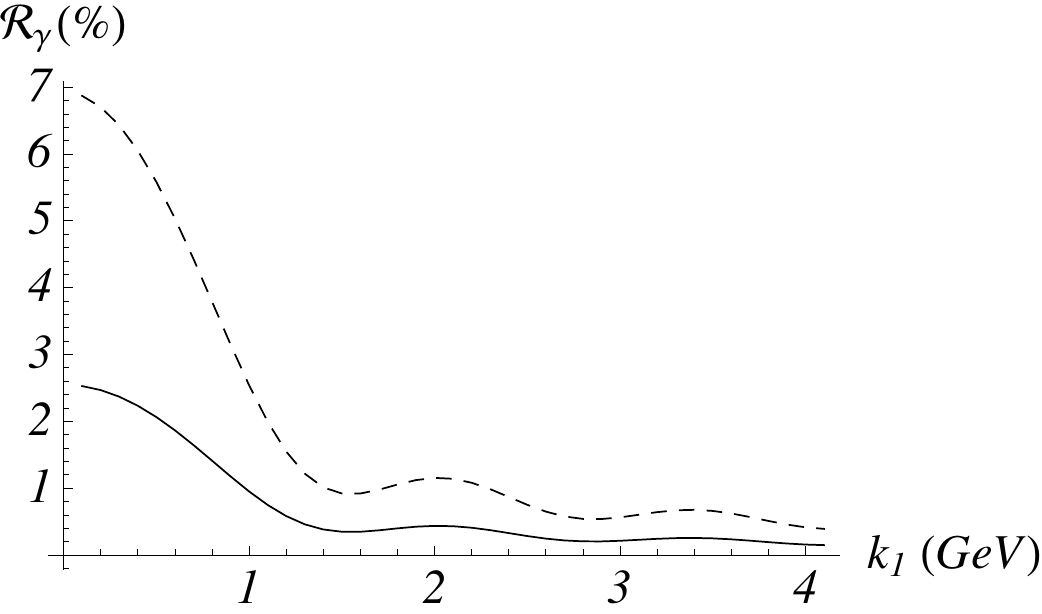} 
 \caption{Fraction of the QED contribution in the total differential photon production cross section. Solid line: Cu, dashed line: Au, $\sqrt{s}= 200$~GeV, $\mu$=1/fm, quark mass $m_q=150$~MeV.}
\label{fig3}
\end{figure}
%%%%%

The relative magnitude of the Coulomb correction to the photon spectrum can be expressed in terms of the ratio
\ball{c55}
\mathcal{R}_\gamma= \frac{d\sigma_\text{em}^{qA\to \gamma q X}}{d^2 k_1 dz}\bigg/ \frac{d\sigma^{qA\to \gamma q X}}{d^2 k_1 dz} \,,
\gal
which is plotted in \fig{fig3}.  As expected, the Coulomb correction is largest at small $k_1$ and for heavier nucleus. For p-Au collisions at $\sqrt{s}=200$~GeV it constitutes about 7\% at small $k_1$. At larger energies it slowly increases as $\log s$.

%%%%%%%%%%%%%%%%%%%%%%%%%%%
\section{Dilepton production}\label{sec:d}

%%%%
%\subsection{Dilepton spectrum at large invariant masses}

Dilepton production by an incident quark is quite complicated because both the quark and the produced dileptons interact with the electromagnetic field of the target nucleus, and quark also interacts with the nuclear color field. At large invariant mass $M$ of produced dilepton pair, an intermediate process of photon splitting $\gamma^*\to \ell^+\ell^-$ can be factored out, which leads to significant simplifications. This will be our assumption throughout this section.  A detailed analysis of this approximation can be found in \cite{Tuchin:2012cd}. 

Our notation scheme in this section follows the same pattern as in the previous one. Momenta of incident photon and outgoing leptons are $q$, $k_1$ and $k_2$ correspondingly; lepton's light-cone momentum fraction is  $z=k_{1+}/q_+$. Transverse coordinates of leptons are $\b x_1$ and $\b x_2$; $\b r= \b x_1-\b x_2$ is dipole size, $\b b= (\b x_1+\b x_2)/2$ its impact parameter. Prime indicates coordinates in the complex conjugated amplitude.  With these notations 
the double inclusive cross section for dilepton production reads
\ball{d11}
\frac{d\sigma ^{\gamma^*A\to  \ell^+\ell^-}}{d^2k_1d^2k_2}= &
\frac{\pi}{(2\pi)^6}\int dz \int d^2x_1 d^2x_2 d^2x_{1}'d^2x_{2}' e^{-i\b k_1\cdot (\b x_1-\b x_{1}')}e^{-i\b k_2\cdot (\b x_2-\b x_{2}')}\phi^{\gamma^*\to  \ell^+\ell^-}(\b r,\b r',z)\nonumber\\
&\times \left[ 1+\mathcal{Q}_\text{em}(\b x_1,\b x_2,\b x_1',\b x_2') - \mathcal{S}_\text{em}(\b b, \b r)  
- \mathcal{S}_\text{em}(\b b', \b r')\right]\,,
\gal
where the squared light-cone wave-function describing photon splitting into dilepton pair  is  given by
\ball{d13}
\phi^{\gamma^*\to  \ell^+\ell^-}(\b r,\b r',z)=&\frac{2\alpha}{\pi}m^2\bigg\{
\frac{\b r\cdot \b r'}{r r'}\, K_1(r m_\ell)K_1(r'm_\ell)
[z^2+(1-z)^2]  + K_0(rm_\ell)K_0(r'm_\ell)   \bigg\}\,.
\gal
 The scattering matrix elements of electric dipole is (cp.\ \eq{c15})
\beql{d15}
\mathcal{S}_\text{em}(\b b, \b r)=\cos\left\{ Z\aver{i\Gamma^{q\bar qN}_\text{em}(\b b,s;\b r)}\right\}  = \cos\left\{\frac{Z}{A}\rho\, 2 \alpha  \int d^2b_a\, T(b_a)\,\ln 
\frac{|\b b-\b b_a-\b r/2|}{|\b b-\b b_a+\b r/2|}\right\}\,,
\eeq
and that of  electric quadrupole is  $\mathcal{Q}_\text{em}$. The later is a complicated function of its coordinates.  Explicit form of its QCD analogue   can be found in \cite{Dominguez:2011wm};  it significantly simplifies in the large $N_c$ approximation \cite{JalilianMarian:2004da}. If either $\b x_1=\b x_{1}'$ or $\b x_2=\b x_{2}'$, the quadrupole reduce to a dipole, e.g.\ 
\ball{d17} 
\mathcal{Q}_\text{em}(\b x_1,\b x_2,\b x_1',\b x_2')|_{\b x_2=\b x_2'} &=\mathcal{S}_\text{em}\big((\b x_1+\b x_1')/2, \b x_1-\b x_1' \big)\,.
\gal
Upon integration over $\b k_1$ and $\b k_2$, eq.~\eq{d11} gives the total inclusive cross section that agrees with results of \cite{Ivanov:1998ka}. 

Since we are interested in invariant mass distribution, it is convenient to introduce another pair of independent momenta, photon transverse momentum $\b q$ and the relative momentum of the pair $\b \ell$, as follows
\ball{d19}
\b q= \b k_1+\b k_2\,,\qquad \b \ell= (1-z)\b k_1-z\b k_2\,.
\gal  
Invariant mass of dilepton can be expressed as
\ball{d21}
M^2= (k_1+k_2)^2= q_+(k_{1-}+k_{2-})-(\b k_1+\b k_2) ^2= \frac{m^2+\b \ell^2}{z(1-z)}\,.
\gal
 We took into account that  in the light-cone perturbation theory $q_-\neq k_{1-}+k_{2-}$ because photon splitting is only an intermediate step in dilepton production.
Using these notations, the phase in \eq{d11} can be written as 
\ball{d37}
-i\b k_1\cdot (\b x_1-\b x_{1}')-i\b k_2\cdot (\b x_2-\b x_{2}')= 
-i\b \ell\cdot (\b r-\b r')-i\b q\cdot(\b b -\b b') -i \b q\cdot (\b r-\b r')  (z-1/2)\,.
\gal
Factorization of photon decay  assumes that  $\ell \sim 1/M$ and $q<2m_\ell $\cite{Tuchin:2012cd}. Therefore, we can neglect the last term in \eq{d37}:
\ball{d38}
-i\b k_1\cdot (\b x_1-\b x_{1}')-i\b k_2\cdot (\b x_2-\b x_{2}')\approx 
-i\b \ell\cdot (\b r-\b r')-i\b q\cdot(\b b -\b b') \,.
\gal
For an almost on-mass-shell photon, the transverse polarization is dominant. Expanding \eq{d13} at small $m_\ell$ and keeping only the term dominant at small dipole sizes,  we get
\ball{d39}
\phi^{\gamma^*\to  \ell^+\ell^-}(\b r,\b r',z)\approx \frac{2\alpha}{\pi}
\frac{\b r\cdot \b r'}{r^2 r'^2}\, 
[z^2+(1-z)^2] \,.
\gal
Since \eq{d39}, as well as the scattering factors are $q$-independent, we can integrate in \eq{d11} over $\b q$, which in view of \eq{d38}, yields $(2\pi)^2\delta(\b b-\b b')$. Moreover, since $M$ is larger than the typical momentum transfer $ \Delta\sim \sqrt{\alpha Z}/b$ by a $t$-channel photon, i.e.\ $\Delta\ll M$,  we can expand the quadrupole amplitude at small difference $|\b r-\b r'|\ll |\b r+\b r'|/2$, which yields $\mathcal{Q}_\text{em}\approx \mathcal{S}_\text{em}(\b b, \b r-\b r')$. (Other scattering factors in \eq{d11} do not depend on this difference). With these assumptions and approximations we derive at large $M$
\ball{d43}
\frac{d\sigma ^{\gamma^*A\to  \ell^+\ell^-}}{d^2\ell d^2b}= &
\frac{\pi}{(2\pi)^4}
\int dz [z^2+(1-z)^2] \int d^2r d^2r'\,  e^{-i\b \ell\cdot(\b r-\b r')}\frac{2\alpha}{\pi}
\frac{\b r\cdot \b r'}{r^2 r'^2}\,\nonumber\\
& \times \left[ 1+\mathcal{S}_\text{em}(\b b, \b r-\b r')- \mathcal{S}_\text{em}(\b b, \b r) -\mathcal{S}_\text{em}(\b b, \b r')  \right]\,.
\gal
We can take one of the two-dimensional integrals using \eq{c25},\eq{c26}.  This gives
\ball{d45}
\frac{d\sigma ^{\gamma^*A\to  \ell^+\ell^-}}{d^2\ell d^2b}= &
\frac{\pi}{(2\pi)^4}\frac{2\alpha}{\pi}\int dz [z^2+(1-z)^2]\frac{2\pi}{\ell^2}
\int d^2r   e^{-i\b \ell\cdot \b r}\ln \frac{1}{r}
\nabla_{\b r}^2 \left[ 1- \mathcal{S}_\text{em}(\b b, \b r)  \right]\,.
\gal
To calculate the Laplacian appearing in the right-hand-side of  \eq{d45} we use the expression for the scattering amplitude in the integrand of \eq{b41} (with $\Gamma_\text{s}=0$):
\ball{d47}
\nabla_{\b r}^2 \left[ 1- \mathcal{S}_\text{em}(\b b, \b r)  \right]=&(2\alpha Z)^2\frac{b^2}{R_A^4}\cos\left( 2\alpha Z \frac{\b b\cdot \b r}{R_A^2}\right)\theta(R_A-b)\nonumber\\
&+\frac{b^2}{(\b b-\b r/2)^2(\b b+\b r/2)^2}(2\alpha Z)^2\cos\left( 2\alpha Z \ln \frac{|\b b-\b r/2|}{|\b b+\b r/2|}\right)
\theta(R_A-b)
\gal
As mentioned before, at $b<R_A$ we can expand this expression in powers of $r^2/R_A^2$, while at $b>R_A$ in powers $r^2/b^2$. We have
\ball{d49}
\nabla_{\b r}^2 \left[ 1- \mathcal{S}_\text{em}(\b b, \b r)  \right]\approx  (2\alpha Z)^2\left[\frac{b^2}{R_A^4}\theta(R_A-b)
+\frac{1}{b^2}
\theta(R_A-b)\right]\,.
\gal
Plugging this into \eq{d45} and employing \eq{c29} yields
\ball{d51}
\frac{d\sigma ^{\gamma^*A\to  \ell^+\ell^-}}{d^2\ell d^2b}= 
\frac{4}{3\pi^2}\frac{\alpha}{\ell^4}(\alpha Z)^2\left[\frac{b^2}{R_A^4}\theta(R_A-b)
+\frac{1}{b^2}
\theta(R_A-b)\right]\,.
\gal
Notice that the dilepton spectrum at a given impact parameter   is energy-independent. This a consequence of the quasi-classical approximation. 
Integration over impact parameter can be done directly in \eq{d45} using \eq{c29} and \eq{b44} if neglect a small contribution at $b<R_A$. The result is
\ball{d58}
\frac{d\sigma ^{\gamma^*A\to  \ell^+\ell^-}}{d^2\ell}&= \frac{\pi}{(2\pi)^4}\frac{2\alpha}{\pi}\int_0^1 dz [z^2+(1-z)^2]\frac{2\pi}{\ell^2}
\int d^2r   e^{-i\b \ell\cdot \b r}\ln \frac{1}{r}
 8\pi (\alpha Z)^2\ln\frac{s}{4m_\ell^2m_NR_A}\nonumber\\
 &=\frac{8\alpha}{3\pi}\frac{1}{\ell^4} (\alpha Z)^2\ln\frac{s}{4m_\ell^2m_NR_A}\,.
\gal
The same formula is obtained by integration of  an approximate formula \eq{d51} over $b$. This is because \eq{b44} assumes that $b\gg R_A$. Note that $b$-integrated cross section 
exhibits logarithmic dependence on energy, which enters through the cutoff $b_\text{max}$ (see \eq{b44}).

If there were no QED interactions of dilepton with the nucleus  we would have instead of \eq{d43}
\ball{d63}
\frac{d\sigma_0 ^{\gamma^*\to  \ell^+\ell^-}}{d^2\ell d^2b}= &
\frac{\pi}{(2\pi)^4}\int dz [z^2+(1-z)^2]\int d^2r d^2r'\,  e^{-i\b \ell\cdot(\b r-\b r')}\frac{2\alpha}{\pi}
\frac{\b r\cdot \b r'}{r^2 r'^2}\nonumber\\
&= \frac{\pi}{(2\pi)^2}\int dz [z^2+(1-z)^2]\frac{2\alpha}{\pi}\frac{1}{\ell^2}= \frac{\alpha}{3\pi^2}\frac{1}{\ell^2}\,,
\gal
Changing the integration variable from $\ell$ to $M$ we obtain the well-known QED result for virtual photon decay probability 
\ball{d65}
\frac{d\sigma_0 ^{\gamma^*\to  \ell^+\ell^-}}{ d^2b}= \frac{2\alpha}{3\pi}\frac{dM}{M}\,.
\gal

The difference between the dilepton production cross section in the Coulomb field and in vacuum can be expressed as the following ratio
\ball{d69}
f(\ell,b)=  \frac{d\sigma ^{\gamma^*A\to  \ell^+\ell^-}}{ d^2b d^2\ell}\bigg/\frac{d\sigma_0 ^{\gamma^*\to  \ell^+\ell^-}}{ d^2\ell d^2b}\,. 
\gal
Using \eq{d51}, \eq{d63} we derive that at large invariant masses 
\ball{d71}
f(\ell,b)= \frac{4(\alpha Z)^2 }{R_A^2\ell^2}\left[\frac{b^2}{R_A^2}\theta(R_A-b)
+\frac{R_A^2}{b^2}
\theta(R_A-b)\right]\,.
\gal

\begin{figure}[ht]
     \includegraphics[width=8cm]{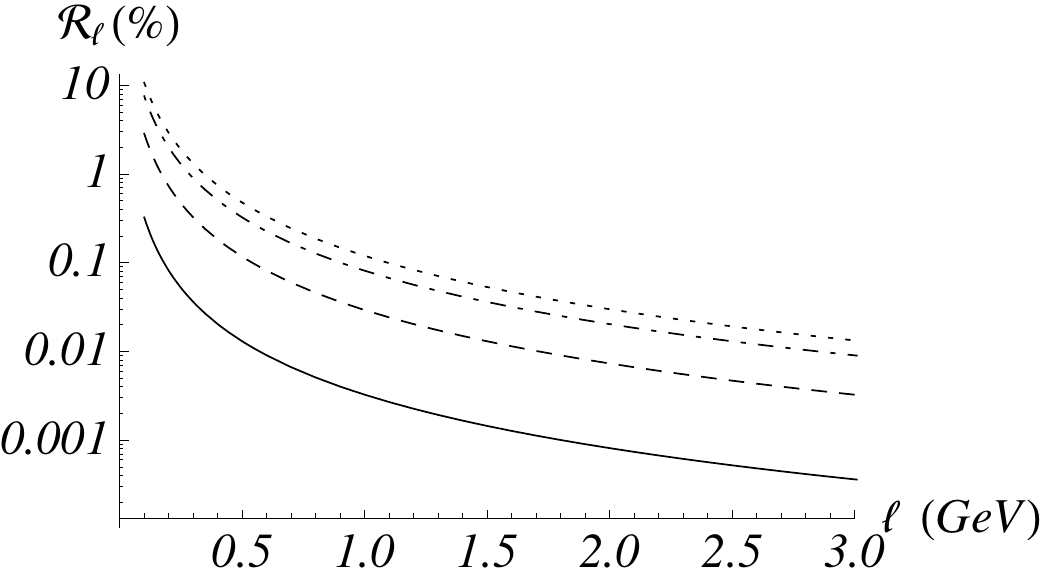} 
 \caption{Fraction of the QED contribution in the $e^+e^-$ dilepton   production cross section in the Coulomb field of gold nucleus, $A=197$, $Z=79$. Solid line: $b=1$~fm, dashed line: $b=3$~fm, dashed-dotted line: $b=5$~fm, dotted line: $b=7$~fm. }
\label{fig4}
\end{figure}

As in the previous section,  we express the relative magnitude of the Coulomb correction to the dilepton spectrum as a  ratio
\ball{d80}
\mathcal{R}_{\ell}= \frac{f_{\ell}}{1+f_{\ell}}\,,
\gal
 which is plotted in \fig{fig4} for electron-positron pair production by high energy  virtual photon in a Coulomb field of gold nucleus. We observe that the relative contribution of the Coulomb corrections to dilepton production increases at smaller $M\sim 2\ell$ and toward the nucleus boundary and can reach 10\% in semi-peripheral and peripheral collisions. 
%%%%
%%%%%

%%%%%%%%%%%%%%%%%%%%%%%%%%%%%%%%%%%%%%%%
\section{Discussion and summary}\label{sec:f}

In this article we investigated the role of electromagnetic Coulomb interactions in photon and dilepton production in high energy pA collisions. Among other important processes that receive electromagnetic corrections is gluon emission off a fast quark and $q\bar q$ production. Photon production vanishes in the eikonal approximation, i.e.\ when valence quark moves strictly along the straight line, corresponding to  $z\to 0$. In contrast, gluon production cross section diverges in this limit as $1/z$ giving the leading logarithmic term to the rapidity distribution.  Therefore, QED contribution to gluon production appears only as a correction to a sub-leading order in $\as$ and can be safely neglected. In $q\bar q$ production via gluon splitting, Coulomb corrections come about already at the leading order because at least one fermion carries finite $z$. 

QED corrections to photon production are largest  at small transverse momentum of photon and increase with energy and nuclear weight. In p-Au collisions at $\sqrt{s}= 200$~GeV per nucleon, the Coulomb correction to photon production reaches 7\%. Dilepton production receives QED contributions at two stages: at virtual photon emission, which is qualitatively similar to photon production, and at virtual photon splitting into a dilepton pair.  The later can proceed even in vacuum. We computed the Coulomb correction to this process and found that it is largest for small invariant masses $M$ and increases with impact parameter. In p-Au collisions at $\sqrt{s}= 200$~GeV per nucleon, the Coulomb correction is up to  10\% at $M\sim 200$~MeV.  An upshot of this is that  the prompt photon yield extracted from the dilepton spectrum using the equation $\frac{dN^{\ell^+\ell^-}}{dM}= \frac{2\alpha}{3\pi M}N^\gamma$, is overestimated by about 10\%.

It is of a special interest to extend the analysis of this article to  the initial stage of heavy-ion collisions. At a qualitative level, we expect that the main features that we observed in pA scattering are carried over to AA scattering. However, a quantitative estimate of the Coulomb corrections in heavy-ion collisions  require further analytical investigation.

%%%%%%%%%%%%%%%%%%%%%%%%%%%%%%%%
\acknowledgments
%I  am grateful to ... for many fruitful discussions of related problems. 
%I thank ...  for useful  correspondence and ... for discussions of related topics. 
This work  was supported in part by the U.S. Department of Energy under Grant No.\ DE-FG02-87ER40371.

%%%%%%%%
\appendix
\section{}\label{appA}

Integral appearing in \eq{b39} can be written in the cylindrical nucleus model as follows
\ball{app1}
I = \int d^2b_a \,\ln \frac{|\b b-\b r/2-\b b_a|}{|\b b+\b r/2-\b b_a|}\,\theta(R_A-b_a)= 
\int d^2b_a \ln \frac{|\b x-\b b_a|}{|\b y-\b b_a|}\,\theta(R_A-b_a)\,,
\gal
where $\theta$ is the step function and we denoted $\b x= \b b -\b r/2$ and $\b y=\b b +\b r/2$. Introducing a dimensionless variable $\xi= b_a/x$ we have 
\ball{app3}
\int d^2b_a \ln |\b x-\b b_a|\theta(R_A-b_a)&= \frac{1}{2}x^2\int_0^{R_A/x}d\xi\, \xi\int _0^{2\pi} d\phi \left[\ln x^2+\ln (1+\xi^2-2\xi \cos\phi)\right]\nonumber\\
&=\pi x^2\int_0^{R_A/x}d\xi\, \xi\left[ \ln x^2+\ln \frac{2}{1+\xi^2+|\xi^2-1|}\right]\nonumber\\
& =\frac{\pi}{2}\times 
\left\{ 
\begin{array}{cl}
x^2-R_A^2+R_A^2\ln R_A^2\,, &  x\ge R_A\,,\\
R_A^2\ln x^2\,,& x<R_A\,.
\end{array}
\right.
\gal
Suppose now for definitiveness that $x>y$. Then 
\ball{app5}
I= \frac{\pi}{2}\times
\left\{ 
\begin{array}{cl}
2 R_A^2\ln \frac{x}{y}\,, &  x,y> R_A\,,\\
 R_A^2\ln \frac{x^2}{R_A^2}+R_A^2-y^2\,, & x>R_A>y\,,\\
x^2-y^2\,, & x,y<R_A
\end{array}
\right.
\gal

%%%%%%%%%%%%%%%%%%%%%%%%%%%%%%%%%%%%%

\end{document}